# Distributed Rank-1 Dictionary Learning:
## Towards Fast and Scalable Solutions for fMRI Big Data Analytics

Milad Makkie*, Xiang Li*, Mojtaba Sedigh Fazli,
Tianming Liu#, Shannon Quinn#

Department of Computer Science,
The University of Georgia, Athens, GA
tianming.liu@gmail.com
*Equal contributions; #Joint corresponding authors.

Binbin Lin, Jieping Ye
Department of Computational Medicine and Bioinformatics
University of Michigan, Ann Arbor, MI

*Abstract*— The use of functional brain imaging for research and diagnosis has benefitted greatly from the recent advancements in neuroimaging technologies, as well as the explosive growth in size and availability of fMRI data. While it has been shown in literature that using multiple and large scale fMRI datasets can improve reproducibility and lead to new discoveries, the computational and informatics systems supporting the analysis and visualization of such fMRI big data are extremely limited and largely under-discussed. We propose to address these shortcomings in this work, based on previous success in using dictionary learning method for functional network decomposition studies on fMRI data. We presented a distributed dictionary learning framework based on rank-1 matrix decomposition with sparseness constraint (D-r1DL framework). The framework was implemented using the Spark distributed computing engine and deployed on three different processing units: an in-house server, in-house high performance clusters, and the Amazon Elastic Compute Cloud (EC2) service. The whole analysis pipeline was integrated with our neuroinformatics system for data management, user input/output, and real-time visualization. Performance and accuracy of D-r1DL on both individual and group-wise fMRI Human Connectome Project (HCP) dataset shows that the proposed framework is highly scalable. The resulting group-wise functional network decompositions are highly accurate, and the fast processing time confirm this claim. In addition, D-r1DL can provide real-time user feedback and results visualization which are vital for large-scale data analysis.

*Keywords:* **fMRI, big data, dictionary learning, functional network, real-time visualization**

## I. INTRODUCTION

In the current field of functional neuroimaging research, one of the most effective approaches for fMRI data analysis is the functional network decomposition based on Dictionary Learning methods [1, 2]. Dictionary learning derives a set of vectors that sparsely code the input fMRI data. The resulting dictionaries and sparseness-constraint loading coefficients respectively characterize the underlying temporal and spatial distribution patterns of the atomic functional networks over the whole brain. Both individual [3] and group-wise dictionary learning studies [1] on fMRI data have been conducted on task and resting-state fMRI data. It has been shown in various studies that specific functional network alterations could help identify brain disorders (e.g. Schizophrenia as in [4], Alzheimer's disease as in [5]), effectively serving as "functional biomarkers" for early diagnosis and potential intervention. Moving forward, functional brain network analysis techniques face major challenges in big data [6]. Firstly, the huge number of available subjects in recent public datasets [7] calls for rapid feedback and interpretation of the analysis results; real-time visualization of the individual functional network would be very important for fast screening and quality control [8]. Secondly, fast and scalable tools for analyzing the large-scale data at group or population level are much in needed. It has been shown that population-level studies on terabytes of data size [7] could offer novel perspectives toward understanding the holistic functional space of human brain and greatly enhance the knowledge discovery from fMRI analysis. Such computational challenges from large-scale data will eventually demand the coordinated resources of distributed clusters [9].

Furthermore, big data analytics tools will not fill the big gap of a comprehensive analytics alone where visualization plays a significant role. An ideal online visualization tool is also necessary for both visually representing the data before, during and after processing as well as understanding the actual method(s)' performance.

Following the previous success in using dictionary learning for functional network decomposition [1, 2], in this work we developed a novel distributed rank-1 dictionary learning (D-r1DL) framework [19], using the Apache Spark platform. All code presented in this work can be found in the Github repository https://github.com/quinngroup/dr1dl-pyspark

Dictionary learning has been in the center of attention of researchers in variety of disciplines, but less effort has been made to create new algorithms using it. In one method [14], the authors designed a large-scale sparse coding framework on Hadoop using the MapReduce distributed programming paradigm. The core operations of dictionary learning were split in two main phases: the sparse coding phase, in which the loading weights were learned in parallel and on different machines (the *map* phase); and the dictionary learning phase, in which the dictionary atoms were updated (the *reduce* phase). By taking advantage of hard sparsity constraints, the authors avoided materializing the entire data matrix in memory at once, instead operating on blocks of the matrix in parallel and constructing the loading matrix row by row. Another recent method was proposed in [18], in which the authors converted dictionary learning into a streaming algorithm. In this way, partial solutions were constructed given only a few rows of input at a time. Their implementation could also handle large dimensions by subsampling the input. However, while this approach was scalable out-of-core on a single machine, it nevertheless precluded datasets which spanned multiple physical machines and for which subsampling was not preferred. Furthermore, it was only as scalable as the single machine on which it was run.

In our proposed method, we have used a similar dataflow technique as [14], but have implemented the core algorithm on the Spark platform instead of Hadoop MapReduce. We assumed the sparse coding and dictionary atoms fit in the memory, which result in overall fast and efficient computation. Where Hadoop MapReduce excels in batch processing, Spark is far more flexible, optimized for iterative computation and capable of parallel computation that requires small communication overhead: intermediate results are cached in-memory on the workers and re-used in subsequent iterations, avoiding the re-serialization and deserialization steps that are unavoidable between Hadoop map and reduce steps. Furthermore, updates in Spark are efficiently broadcasted to the workers. Most importantly, the low-level Spark distributed data abstraction, the Resilient Distributed Dataset (RDD), pipelines the requested transformation operations and lazily executes them after an action is issued, using a constructed DAG of transformations to determine the optimal computational pathway. We leverage these advantages to provide a substantial performance gain in our D-r1DL dictionary learning implementation.

The performance of the proposed framework on both individual and group-wise data from Human Connectome Project (HCP) Q1 release [10] shown that it is a suitable solution for fMRI big data analytics. Although the data size itself might not directly address the big-data challenge but using the generated data matrix, we have shown that the total processing time of the same dataset can be reduced by half in comparison to the stochastic coding dictionary learning algorithm without the parallelization. This by itself shows that we have achieved a method to meet the challenge posed by fMRI big data for more efficient and scalable data analytics method.

The goal is to show how our new algorithm will affect the total processing time and memory usage. The two important constrains that the neuroimaging community has faced with in the past few years. While the data-sets size is growing exponentially, the computational architecture is not capable of handling such a big growth.

The main features of our proposed D-r1DL are as follow:

1) **Accuracy**: D-r1DL can discover the same set of results by the General Linear Model (GLM), as well as other functional networks reported in literature such as the well-known resting-state networks (RSNs) also from task-based (tfMRI) data.
2) **Speed**: D-r1DL distributes computational loads to many nodes, thus achieving greater speed increases with larger clusters. On individual data, the decomposition results are visualized and fed to the user in real-time.
3) **Scalability**: D-r1DL has near-constant memory cost regardless of the input data size as the nodes work on partitions of data rather than the whole dataset. Spark's basic distributed data abstraction, the resilient distributed dataset (RDD), is designed to scale gracefully with the size of the data. In addition, the memory cost of the learning process is minimized to only two vectors.
4) **Deployment:** D-r1DL has been integrated into our in-house neuroinformatics system, HELPNI (HAFNI-enabled largescale platform for neuroimaging informatics), as introduced in [17] publicly available at http://bd.hafni.cs.uga.edu.

## II. MATERIALS AND METHODS

### A. Algorithm of rank-1 matrix decomposition with sparse constraint

The rank-1 dictionary learning (r1DL) algorithm decomposes the input matrix $S$ (of dimension $T \times P$) by iteratively estimating the basis vector $u$ ($T \times 1$ vector with unit length) and the loading coefficient vector $v$ ($P \times 1$ vector). The algorithm is an extreme case of the general dictionary learning framework [11] as the input is approximated by a rank-1 matrix (spanned by two vectors).

With the *l*-0 sparseness constraint, the following energy function $L(u, v)$ will be minimized:

$$L(u, v) = \|S - uv^T\|_F, \text{ s.t. } \|u\| = 1, \|v\|_0 \leq r. \quad (1)$$

Thus the total number of non-zero elements in $v$ should be smaller than or equal to the given sparsity constraint parameter $r$ which is generally empirically determined based on the context of the application. The algorithm alternates updating $u$ (randomly initialized before the first iteration) and $v$ until the convergence of $u$:

$$v = \underset{v}{\operatorname{argmin}} \|S - uv^T\|_F, \text{ s.t. } \|v\|_0 \leq r,$$
$$u = \underset{u}{\operatorname{argmin}} \|S - uv^T\|_F = \frac{Sv}{\|Sv\|} \quad (2)$$

One dictionary basis $[u, v]$ can be estimated after the convergence of Eq. 2. Because the value of the energy function in Eq. 1 decreases at each iteration in Eq. 2, the objective function is guaranteed to converge. For estimating the next dictionary (up to the dictionary size $K$), the input matrix $S$ will be deflated to its residual $R$.

$$R^n = R^{n-1} - v^T R^{n-1}, R^0 = S, 1 < n \leq K. \quad (3)$$

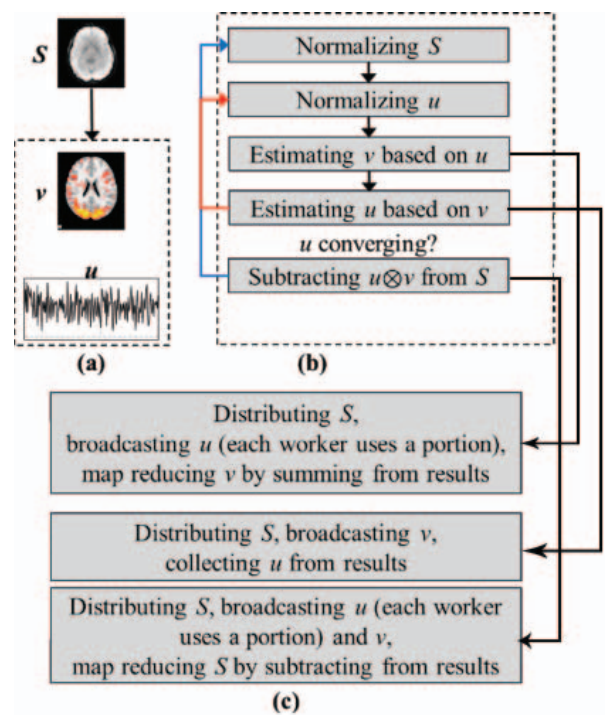

Figure 1. Illustration of the D-r1DL framework. (a) Running example showing the input data S (one volume from the 4-D volumetric matrix), learned vector v (3-D volumetric matrix as a vector) and vector u (time series). (b) Algorithmic pipeline of r1DL. Red arrow shows the updating loop for learning each [u, v], blue arrow shows the updating loop for deflation of S and learning next dictionary. (c) Parallelization steps for the three operations from (b).

*All codes are accible in the Github repository https://github.com/quinngroup/dr1dl-pyspar

### B. Distributed dictionary learning framework supported by Spark

In order to utilize computational power and memory capacity across many machines to address the big data problem, we implemented the distributed r1DL algorithm on Spark, which we refer to as the distributed rank-1 dictionary learning (D-r1DL) framework as illustrated in Fig. 1. Using Spark's Resilient Distributed Dataset (RDD) abstraction from [12], D-r1DL can potentially deal with large-scale imaging data whose size exceeds the memory capacity of the working machine. Spark addresses such out-of-core operations by loading only specific partitions of the whole input matrix $S$ into the memory of each node. The learning of dictionary bases $[u, v]$ is performed in parallel at each node (i.e. machine), and are then broadcasted across all nodes during the update. Specifically, the matrix multiplication operations described in Eq. 2 and the deflation operation defined in Eq. 3 were implemented by their corresponding distributed primitives in Spark:

I. During the vector-matrix multiplication, each node will use its portion of the updated $u$ vector, then estimate the $v$ vector based on the multiplication of its partition of $S$ and the vector $u$. The resulting partial $v$ vectors from all the nodes will be then reduced by the summation operation.

II. During the matrix-vector multiplication, each node will use the updated $v$ vector and its partition of the $S$ matrix to estimate a single corresponding element of the $u$ vector. The resulting $u$ vector is assembled from the results of each node.

III. During the matrix deflation operation, both $u$ and $v$ learned from Eq. 2 will be broadcasted. Each node estimates a portion of the outer product between corresponding elements of the $u$ vector with the whole $v$ vector. Each partition of the $S$ matrix is deflated using the corresponding partial product of $u$ and $v$.

### C. D-r1DL as a web service with integrated neuroinformatics system

In addition to the open-source code implementing the D-r1DL framework, in this work we proposed an in-house neuroinformatics platform, HELPNI, to publicly host the D-r1DL framework, aiming to perform functional network decomposition analysis as a web service available to collaborators and other researchers. The neuroinformatics platform consists of three main components: the data storage core, data management and web application core, and the data processing core.

1) **Data storage:** Both local hard drives and cloud storage are integrated in the system, as we are using Amazon Simple Storage Solution (S3) as permanent data storage for larger datasets. Data are accessible through this web application with credentials using Postgresql. Users are able to either manually upload the data to the system using the web-based Java uploader, or they can use the script uploading method which allows them to upload theoretically unlimited number of images after defining the appropriate data schema on the system using the bulk script uploading feature. This helps the collaborators to import their desired data-sets hosted in a medical facility through the PACS protocol. Visualizing the stored data can be done in two different ways. If the stored data are in a PACS friendly format, such as DICOM, then it can be easily visualized through a Java plugin in a web browser. But if the data are in other formats, including Nifti, it should be either visualized locally or, alternatively, using the converted DICOM format. The stored fMRI images can carry the standard parameters such as scan type, repetition time, dimension, etc. Users can set them at feeding step or they can be automatically retrieved from DICOM headers or through meta data.

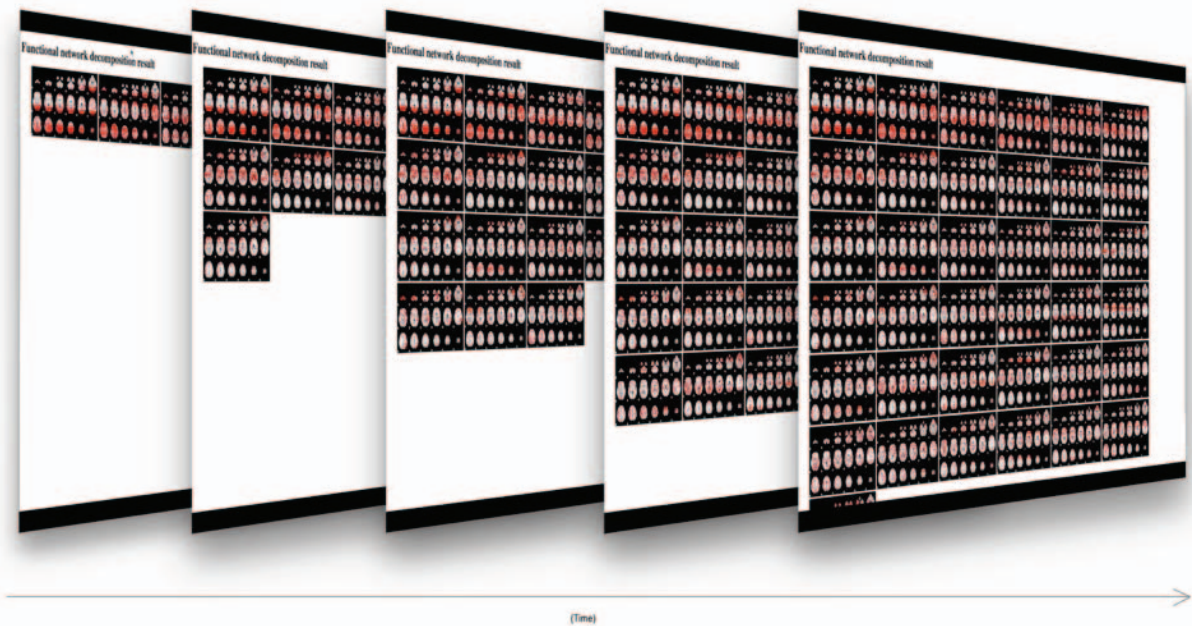

Figure2. The generated networks as being computed will appear on a dynamically-generated result screen linked to the report webpage.

2) **Data management and web application**: We used Apache Tomcat as the web server to handle incoming requests. User-uploaded data is stored in local storage accessible from preparation units. Defining processing pipelines over the selected group of subjects is one of the key features of our neuroinformatics platform, HELPNI, where all the algorithms and processing procedures can be easily assigned to any project using the XML descriptors. This includes the preprocessing and the r1DL algorithms also. For fMRI data, the preparation steps include preprocessing and the conversion of the 4D fMRI images to 2D data matrix. Model parameters are also set during the preparation: either automatically extracted from the data (e.g. $T$ and $P$) or defined by user specification (e.g. sparseness constraint r). While the data are being processed, an online visualization tool will simultaneously generate the reports of the statistics and visualizations of the decomposed functional networks. Fig 2 shows an overview of real time visualization of discovered networks. Then the results will be uploaded to the Apache server, accessible via web browsers for viewing and sharing. The PDF version of all reports as well as a interactive webpage, will be available in every subjects' profile page. This will make the future comparison and studies much easier. Also, all the results will remain at a system directory addressed in the subjects' profile. Doing so will help collaborators future studies be done easier and more efficient. Because they can access raw data as well as any prior study results instantly. For example, the standard fMRI preprocessing can be done one time and all the future analysis can easily leverage from the one time preprocessed data.

3) **Data processing:** HELPNI platform controls the data flow and working schedule from the prepared data to the processing units. One advantage of the proposed neuroinformatics platform is its flexibility with respect to the processing control. Different computational nodes can be used to do the computation simultaneously over different subjects. In this work, the platform controls 3 data processing units: the first one is the in-house cluster (8 cores, 16 GB memory) deployed on the same machine as the platform; the second one is a remote high performance computing cluster (48 cores, 128 GB memory); the third one is the cloud-based Amazon EC2 cluster. Fig. 3 shows an overview of the neuroinformatics system, through which stored fMRI data in centralized storage will be sent to processing units and the results will be visualized through dynamically-generated webpages.

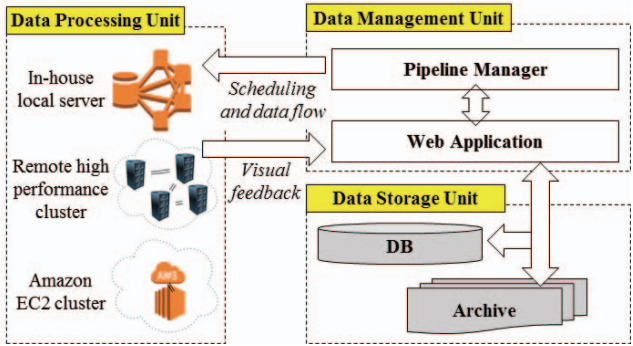

Figure 3. Overview of the proposed neuroinformatics system with the three central units (Data processing, Management and Storage) and their relationships.

## III. EXPERIMENTAL RESULTS

### A. Model performance on a relatively large-scale dataset

We applied the D-r1DL model on the publicly available dataset from Human Connectome Project [10] for validating its effectiveness in discovering functional networks from large-scale fMRI dataset. The acquisition parameters of the fMRI are as follows: 90×104 matrix, 220mm FOV, 72 slices, TR=0.72s, TE=33.1ms, flip angle=52°, BW=2290 Hz/Px, 2.0mm isotropic voxels. Data preprocessing followed the protocols detailed in [13], including motion correction, spatial smoothing, temporal pre-whitening, slice time correction, and global drift removal. The tfMRI data was then registered to the standard MNI152 2mm space using FSL FLIRT to enable group-wise analysis. The final individual tfMRI signal matrix used as model input contains 223,945 number of voxels (defined on grey matter) and varying temporal length based on task design. In this work, tfMRI datasets from 68 subjects during Emotion Processing task were used, with the time length of 176 volumes which matches the aim of the proposed framework for population-level fMRI big data analysis.

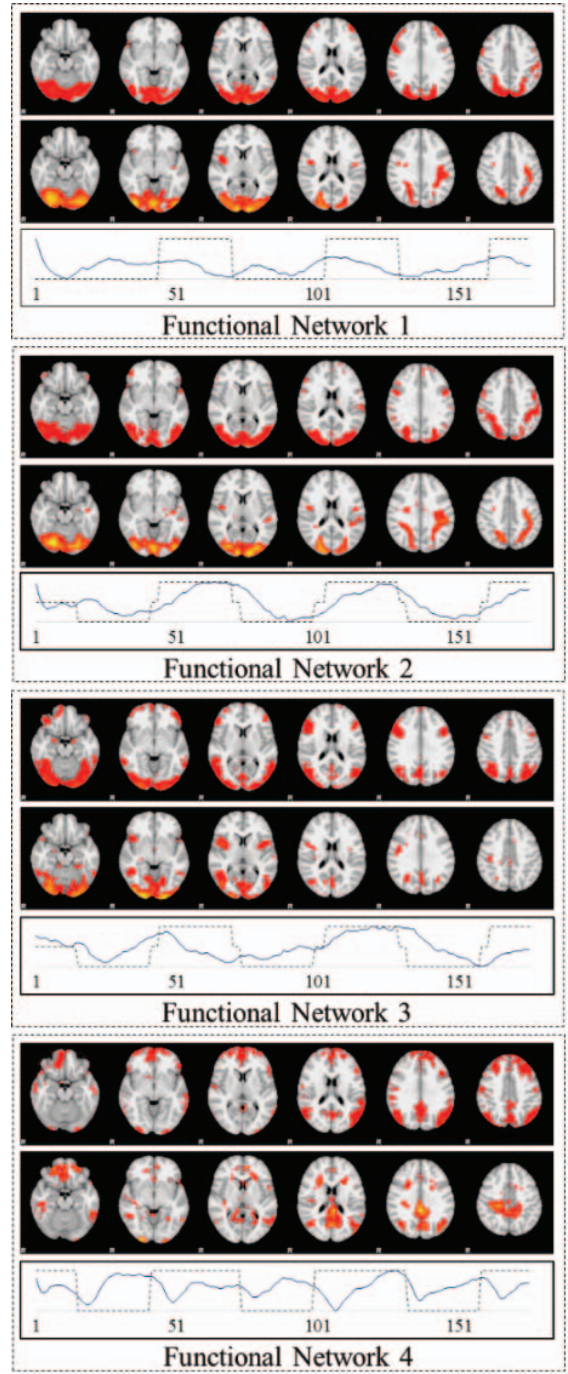

Figure4. Spatial maps of the four pairs of group-wise functional networks obtained by r1DL (upper) and GLM (lower) from Emotion dataset. The temporal pattern of the functional networks are shown below the spatial patterns.

Afterwards, we aggregated the 68 individual fMRI data during Emotion task into one big, group-wise matrix with the dimension of 176×15,228,260 (~20 GB as a text file). Using the parameter setting of $K$=100 (i.e. decomposing 100 functional networks) and $r$=0.07 (i.e. 7% of the total number of grey matter voxels across all subjects can have non-zero value), we obtained the 100 group-wise functional networks. The analysis was performed on the high performance computing cluster and took around 10 hours to finish. The temporal patterns of the group-wise functional networks are defined in the $D$ matrix. The spatial patterns were distributed across each individual's space (223,945 voxels) in the $z$ matrix. To obtain a volumetric image, we averaged the loading coefficient value on each voxel across all individuals.

For validation purposes, we compared the decomposed group-wise functional networks with the group-wise activation detection results obtained by model-driven General Linear Model (GLM). The basic rationale of such comparison is that the activation detection results characterize the intrinsic and basic temporal/spatial patterns as a response to external stimuli and should therefore also be revealed by data-driven matrix decomposition-based methods such as D-r1DL. In order to identify the correspondence between the 100 functional networks decomposed by D-r1DL and the GLM results, we calculated Pearson's correlation between the temporal patterns (in the $D$ matrix) of the functional networks and the hemodynamic response function (HRF)-convolved task designs of Emotion Processing task and selected the result with the highest correlation. The group-wise functional network obtained by D-r1DL and the corresponding GLM results are shown in Fig. 4. We also calculated the spatial overlapping rate $SOR$ between the spatial patterns of the results from D-r1DL ($P_1$) and group-wise GLM ($P_2$) to quantitatively measure their similarity:

$$SOR(P_1, P_2) = |P_1 \cap P_2|/|P_2|, \qquad (4)$$

where operator |•| counts the total number of voxels with non-zero values in the given spatial pattern. The rate ranges from 0 (no voxels overlapping) to 1 (exact the same pattern with GLM result). The $SOR$ values of the four pairs of correspondent results between D-r1DL and GLM are 0.72, 0.75, 0.67 and 0.65, respectively.

### B. Model application with sampling strategy

In addition to the analysis on the whole group-wise tfMRI dataset, we also uniformly sampled the 176×15,228,260 input matrix into 10%~90% of its size (e.g. 10% sampled data is a 176×1,522,826 matrix). The main rationale for the sampling study is to further accelerate initial investigations into the effectiveness of the dictionary bases learned by D-r1DL. In such circumstances, the sampling strategy could offer an approximation of the more detailed and accurate functional networks learned from the original data. By applying D-r1DL on the 9 sampled datasets, the corresponding sets of functional networks were obtained. One example functional network showing the correspondence between the 10 sets of results is visualized in Fig. 5.

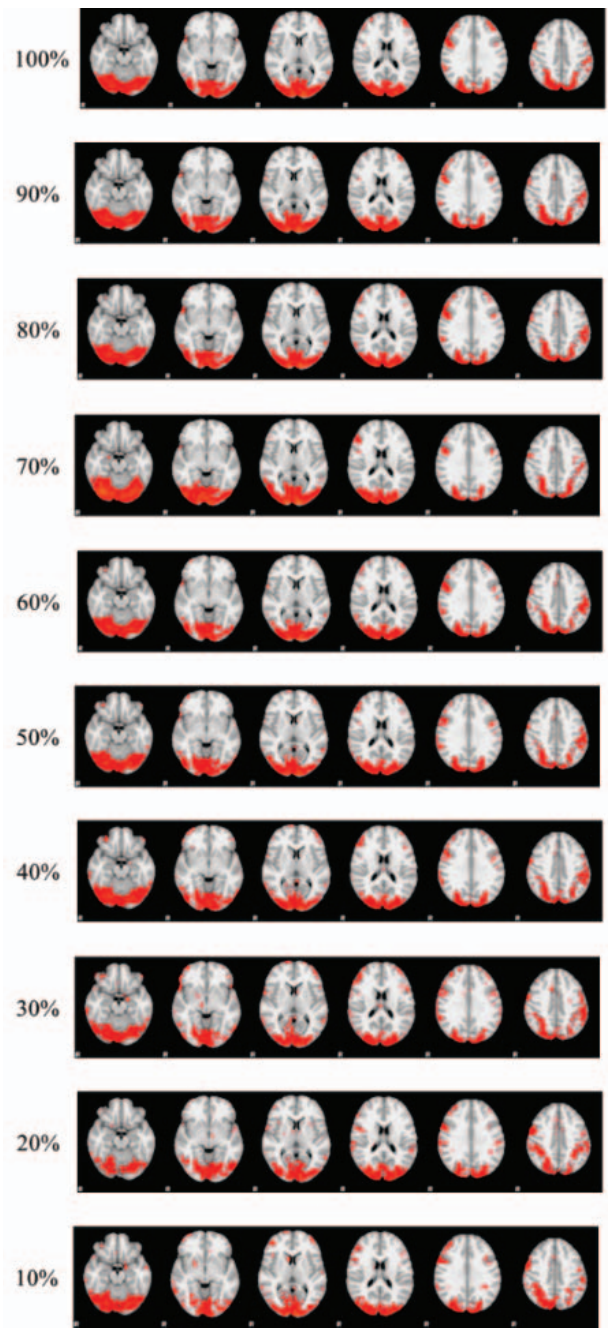

Figure5. Visualization of the spatial patterns of a sample functional networks learned from group-wise aggregated fMRI data with different sampling rates.

It was observed that the spatial patterns of the corresponding functional networks learned from the same dataset with different sampling rates are largely the same (with overlapping rate>0.85), excepting some minor differences in the details. The time costs for the group-wise analysis on uniformly-sampled datasets are summarized in Fig. 6. The time cost follows a quadratic function with the sampling rate ($R^2$=0.994). Thus, while analyzing the original 20 GB dataset took around 10 hours to finish, the time cost is approximately 1 hour using the 20% sampled data.

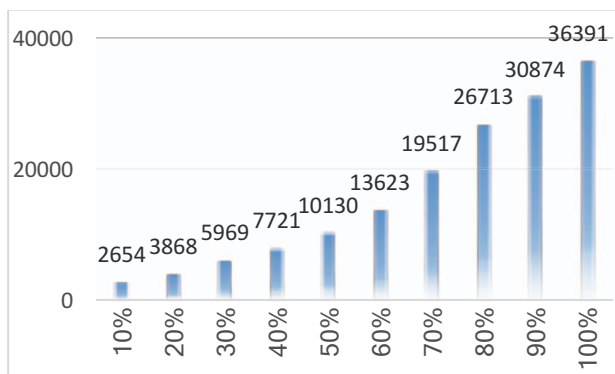

Figure6. Time cost (measured in seconds) for decomposing 100 functional networks from group-wise aggregated fMRI data with different sampling rates. The original dataset has the sampling rate of 100% (rightmost).

### C. Performance boost relative to other dictionary learning algorithms

The advantages of the proposed D-r1DL algorithm are predicated on its smaller memory footprint and robust learning mechanism (no need to set learning rate); even without parallelization, the algorithm should have similar or faster running speed compared with other dictionary learning methods, as Spark intrinsically performs out-of-core computations whether these are distributed over multiple machines or run in parallel on a single machine. We compare D-r1DL with two other dictionary learning algorithms: the online dictionary learning framework implemented in SPAMS [15] and the stochastic coordinate coding (SCC) algorithm introduced in [16]. We applied these two methods on the same HCP Q1 dataset and computed performance statistics compared to D-r1DL. We ran these algorithms using the same in-house server. The performance comparison is shown in Fig. 7 (averaged across all 68 subjects over the HCP task fMRI (tfMRI) dataset). From the comparison, it can be seen that D-r1DL outperformed the other two methods in all the 7 tfMRI datasets.

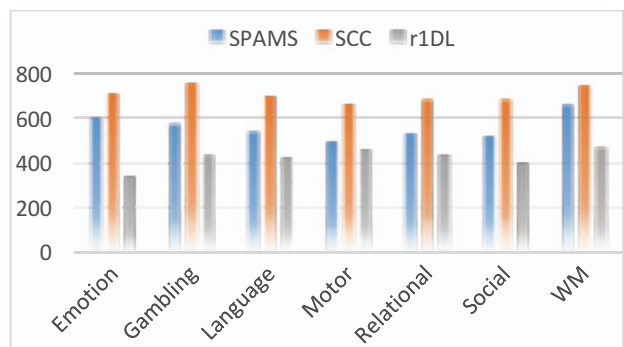

Figure7. Average time cost (measured in seconds) for functional network decomposition from individual tfMRI data during 7 tasks across 68 subjects, using the three dictionary learning methods.

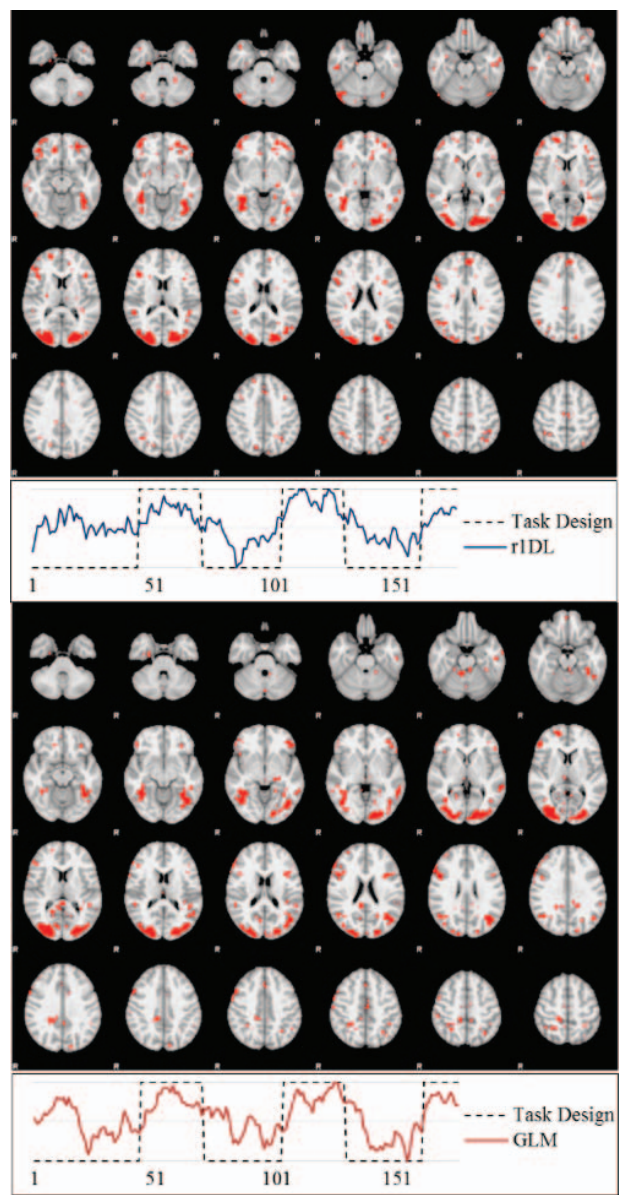

Figure8. Spatial maps and temporal variation patterns of the functional networks decomposed by D-r1DL (left) and GLM (right) on the tfMRI data during Emotion Processing task from a randomly-selected subject.

## D. Real-time user feedback using web-based D-r1DL

We tested the performance of D-r1DL on the neuroinformatics platform as introduced in section 2.3 for individual-level analysis. Using individual fMRI matrix (with dimensions 176×223,945) as input and the same parameter setting as for group-wise analysis ($K$=100, $r$=0.07), the combined time cost for decomposing one network, generating network visualizations, and reporting web pages averaged around 4 seconds on our in-house server. Such a time cost is short enough for real-time visualizations on the decomposition results, thereby providing a useful feedback mechanism for the users. One sample result from the individual-level analysis and the comparison with GLM activation detection results is shown in Fig. 8.

## IV. CONCLUSION AND DISCUSSION

In this work we developed and implemented the D-r1DL framework on Spark for distributed functional network analysis on large-scale neuroimaging data, as well as an online visualization tool, and tested its performance on both the individual and group-wise fMRI data from HCP Q1 release dataset. The results show that the framework can meet the desired scalability and reproducibility requirements for fMRI big data analysis and serve as a useful tool for the community. The framework and the neuroinformatics system are both online as web service for public usage and testing. In addition, we are aiming to provide a general solution for all types of large-scale biomedical/biological imaging data based on D-r1DL. Currently we are working on applying the same algorithm using the Apache Flink framework. While Spark is vastly superior to Hadoop MapReduce for highly iterative computations, Flink possesses a few domain-specific advantages over Spark that could yield additional performance gains for D-r1DL. These include its real-time data streaming engine, instead of Spark's fixed-window batch processing, may boost our current achieved processing speed. Furthermore, while Spark's distributed operations are split between lazy (transformations) and eager (actions), all of Flink's distributed primitives are lazily evaluated. This allows Flink to build and optimize the full dataflow graph before any computation is performed, theoretically minimizing communication overhead of intermediate results to the fullest possible extent. We are also aiming to apply our new algorithm on variety of different neuroimaging data sources including EEG.

References:

[1] Lee, Y.-B., Lee, J., Tak, S., Lee, K., Na, D.L., Seo, S.W., Jeong, Y., Ye, J.C.: Sparse SPM: Group Sparse-dictionary learning in SPM framework for resting-state functional connectivity MRI analysis. NeuroImage 125, 1032-1045 (2016)

[2] Kangjoo, L., Sungho, T., Jong Chul, Y.: A Data-Driven Sparse GLM for fMRI Analysis Using Sparse Dictionary Learning With MDL Criterion. Medical Imaging, IEEE Transactions on 30, 1076-1089 (2011)

[3] Lv, J., Jiang, X., Li, X., Zhu, D., Chen, H., Zhang, T., Zhang, S., Hu, X., Han, J., Huang, H., Zhang, J., Guo, L., Liu, T.: Sparse representation of whole-brain fMRI signals for identification of functional networks. Medical Image Analysis 20, 112-134 (2015)

[4] Calhoun VD, Maciejewski PK, Pearlson GD, KA, K.: Temporal lobe and "default" hemodynamic brain modes discriminate between schizophrenia and bipolar disorder. Hum Brain Mapp. 29, 1265-1275 (2008)

[5] Sorg, C., Riedl, V., Mühlau, M., Calhoun, V.D., Eichele, T., Läer, L., Drzezga, A., Förstl, H., Kurz, A., Zimmer, C., Wohlschläger, A.M.: Selective changes of resting-state networks in individuals at risk for Alzheimer's disease. PNAS, 104, 18760-18765 (2007)

[6] Van Horn, J.D., Toga, A.W.: Human neuroimaging as a "Big Data" science. Brain Imaging and Behavior 8, 323-331 (2013)

[7] Smith, S.M., Hyvärinen, A., Varoquaux, G., Miller, K.L., Beckmann, C.F.: Group-PCA for very large fMRI datasets. NeuroImage 101, 738-749 (2014)

[8] Fan, J., Han, F., Liu, H.: Challenges of Big Data analysis. National Science Review (2014)

[9] Freeman, J., Vladimirov, N., Kawashima, T., Mu, Y., Sofroniew, N.J., Bennett, D.V., Rosen, J., Yang, C.-T., Looger, L.L., Ahrens, M.B.: Mapping brain activity at scale with cluster computing. Nat Meth 11, 941-950 (2014)

[10] Van Essen, D.C., Smith, S.M., Barch, D.M., Behrens, T.E.J., Yacoub, E., Ugurbil, K.: The WU-Minn Human Connectome Project: An overview. NeuroImage 80, 62-79 (2013)

[11] Lee, H., Battle, A., Raina, R., Ng, A.Y.: Efficient sparse coding algorithms. In: Advances in Neural Information Processing Systems, (2006)

[12] Zaharia, M., Chowdhury, M., Das, T., Dave, A., Ma, J., McCauley, M., Franklin, M.J., Shenker, S., Stoica, I.: Resilient distributed datasets: a fault-tolerant abstraction for in-memory cluster computing. Proceedings of the 9th USENIX conference on Networked Systems Design and Implementation, pp. 2-2. USENIX Association, San Jose, CA (2012)

[13] Barch, D.M., Burgess, G.C., Harms, M.P., Petersen, S.E., Schlaggar, B.L., Corbetta, M., Glasser, M.F., Curtiss, S., Dixit, S., Feldt, C., Nolan, D., Bryant, E., Hartley, T., Footer, O., Bjork, J.M., Poldrack, R., Smith, S., Johansen-Berg, H., Snyder, A.Z., Van Essen, D.C.: Function in the human connectome: Task-fMRI and individual differences in behavior. NeuroImage 80, 169-189 (2013)

[14] Sindhwani, V. and Ghoting, A., 2012. Large-scale distributed non-negative sparse coding and sparse dictionary learning. In Proceedings of the Proceedings of the 18th ACM SIGKDD international conference on Knowledge discovery and data mining 2339610, 489-497.

[15] Mairal, J., Bach, F., Ponce, J., and Sapiro, G., 2010. Online Learning for Matrix Factorization and Sparse Coding. J. Mach. Learn. Res. 11, 19-60.

[16] Lin, B., Li, Q., Sun, Q., Lai, M.-J., Davidson, I., Fan, W., and Ye, J., 2014. Stochastic Coordinate Coding and Its Application for Drosophila Gene Expression Pattern Annotation.

[17] Makkie, M., Zhao, S., Jiang, X., Lv, J., Zhao, Y., Ge, B., Li, X. Han, J. and Liu, T. (2015). HAFNI-enabled largescale platform for neuroimaging informatics (HELPNI).Brain informatics, 2(4), 225-238.

[18] Mensch, A., Mairal, J., Thirion, B., & Varoquaux, G. (2016). Dictionary Learning for Massive Matrix Factorization. arXiv preprint arXiv:1605.00937.

[19] Li, X., Makkie, M., Lin, B., Fazli, M. S., Davidson, I., Ye, J., ... & Quinn, S. (2016). Scalable Fast Rank-1 Dictionary Learning for fMRI Big Data Analysis.